\documentclass[11pt,preprint]{aastex}
\usepackage{natbib}
\usepackage{xspace}
\usepackage{xcolor}
\usepackage[
  pdfstartview=FitH,
  linkcolor=blue,
  anchorcolor=blue,
  citecolor=blue,
  urlcolor=blue,
  colorlinks=true,
  breaklinks]{hyperref}
\newcommand{\hili}[1]{\textcolor{black}{#1}\xspace}
\begin{document}

\shorttitle{GG Tau}
\shortauthors{Andrews et al.}

\title{Resolved Multifrequency Radio Observations of GG Tau}

\author{
Sean M. Andrews\altaffilmark{1},
Claire J. Chandler\altaffilmark{2},
Andrea Isella\altaffilmark{3},
T.~Birnstiel\altaffilmark{1},
K.~A.~Rosenfeld\altaffilmark{1},
D.~J.~Wilner\altaffilmark{1},
L.~M.~P{\'e}rez\altaffilmark{2},
L.~Ricci\altaffilmark{3},
J.~M.~Carpenter\altaffilmark{3}, 
N.~Calvet\altaffilmark{4},
S.~A.~Corder\altaffilmark{5}, \\
A.~T.~Deller\altaffilmark{6},
C.~P.~Dullemond\altaffilmark{7},
J.~S.~Greaves\altaffilmark{8},
R.~J.~Harris\altaffilmark{9},
Th.~Henning\altaffilmark{10}, 
W.~Kwon\altaffilmark{11}, \\
J.~Lazio\altaffilmark{12}, 
H.~Linz\altaffilmark{10},
L.~G.~Mundy\altaffilmark{13},
A.~I.~Sargent\altaffilmark{3},
S.~Storm\altaffilmark{13},
L.~Testi\altaffilmark{14,15}
}

\altaffiltext{1}{Harvard-Smithsonian Center for Astrophysics, 60 Garden Street, Cambridge, MA 02138, USA}
\altaffiltext{2}{National Radio Astronomy Observatory, P.O. Box O, Socorro, NM 87801, USA}
\altaffiltext{3}{California Institute of Technology, 1200 East California Blvd., Pasadena, CA 91125, USA}
\altaffiltext{4}{Department of Astronomy, University of Michigan, 500 Church Street, Ann Arbor, MI 48109, USA}
\altaffiltext{5}{Joint ALMA Observatory, Avenida Alonso de C{\'o}rdova 3107, Vitacura, Santiago, Chile}
\altaffiltext{6}{The Netherlands Institute for Radio Astronomy (ASTRON), 7990-AA Dwingeloo, Netherlands}
\altaffiltext{7}{Heidelberg University, Center for Astronomy, Albert Ueberle Str 2, Heidelberg, Germany}
\altaffiltext{8}{University of St.~Andrews, Physics and Astronomy, North Haugh, St.~Andrews KY16 9SS, UK}
\altaffiltext{9}{Department of Astronony, University of Illinois, Urbana, IL 61810, USA}
\altaffiltext{10}{Max Planck Institut f{\"u}r Astronomie, K{\"o}nigstuhl 17, 69117 Heidelberg, Germany}
\altaffiltext{11}{SRON Netherlands Institute for Space Research, Landleven 12, 9747 AD Groningen, The Netherlands}
\altaffiltext{12}{Jet Propulsion Laboratory, California Institute of Technology, 4800 Oak Grove Dr., Pasadena, CA 91106, USA}
\altaffiltext{13}{Department of Astronomy, University of Maryland, College Park, MD 20742, USA}
\altaffiltext{14}{European Southern Observatory, Karl Schwarzschild Str 2, 85748, Garching, Germany}
\altaffiltext{15}{INAF-Osservatorio Astrofisico di Arcetri, Largo E.~Fermi 5, 50125 Firenze, Italy}

\email{sandrews@cfa.harvard.edu}

\begin{abstract}
We present sub-arcsecond resolution observations of continuum emission 
associated with the GG Tau quadruple star system at wavelengths of 1.3, 2.8, 
7.3, and 50\,mm.  These data confirm that the GG Tau A binary is encircled by a 
circumbinary ring at a radius of 235\,AU with a FWHM width of $\sim$60\,AU.  We 
find no clear evidence for a radial gradient in the spectral shape of the ring, 
suggesting that the particle size distribution is spatially homogeneous 
\hili{on angular scales $\gtrsim$0\farcs1}.  A central point source, likely 
associated with the primary component (GG Tau Aa), exhibits a composite 
spectrum from dust and free-free emission.  Faint emission at 7.3\,mm is 
observed toward the low-mass star GG Tau Ba, although its origin remains 
uncertain.  Using these measurements of the resolved, multifrequency emission 
structure of the GG Tau A system, models of the far-infrared to radio spectrum 
are developed to place constraints on the grain size distribution and dust mass 
in the circumbinary ring.  The non-negligible curvature present in the ring 
spectrum implies a maximum particle size of 1--10\,mm, although we are unable 
to place strong constraints on the distribution shape.  The corresponding dust 
mass is 30--300\,$M_{\oplus}$, at a temperature of 20--30\,K.  We discuss how 
this significant concentration of relatively large particles in a narrow ring 
at a large radius might be produced in a local region of higher gas pressures 
(i.e., a particle ``trap") located near the inner edge of the circumbinary 
disk.
\end{abstract}

\section{Introduction} 

The first step of planet formation --- the collisional growth of $\mu$m-sized 
dust grains into $>$km-sized planetesimals, the building blocks of terrestrial 
planets and the cores of giant planets --- is fundamental, but physically 
complicated and fraught with theoretical uncertainty.  A substantial effort 
with numerical simulations and laboratory experiments is converging on a basic 
model framework for the growth and migration of solids embedded in a 
protoplanetary gas disk \citep[see the recent reviews by][]{testi14,
johansen14}, but direct astronomical observations of these solids are required 
to test and refine it.  Thermal continuum emission at mm/radio wavelengths is 
well-suited for that task, as a (relatively) bright and optically thin tracer 
of solid particles with sizes up to $\sim$10\,cm.  The spectral behavior of 
this emission is diagnostic of the particle size distribution 
\citep[e.g.,][]{beckwith91,miyake93,henning96,dalessio01,draine06,ricci10a,
ricci10b}.  Therefore, spatially resolved measurements of the mm/radio 
``colors" can be used to map out how the particle growth and transport 
efficiencies vary as a function of the local physical conditions in the gas 
disk \citep{isella10,banzatti11,guilloteau11,perez12,trotta13,menu14}.  

These preliminary studies of the resolved multifrequency continuum emission 
from disks indicate that the inward radial transport of mm/cm-sized solids is a 
crucial factor for explaining the observed color gradients 
\citep[cf.,][]{birnstiel12}.  \citet{birnstiel14} suggested that this same {\it 
radial drift}, induced by aerodynamic drag on particles that are partially 
coupled to the gas in its sub-Keplerian velocity field \citep{adachi76,
weidenschilling77}, is also responsible for the observed discrepancies between 
the sizes of the line and continuum emission in some disks 
\citep[e.g.,][]{panic09,andrews12,rosenfeld13b}.  However, there is a 
fundamental issue with the transport timescales: in these idealized models, 
radial drift is much too efficient \citep{takeuchi02,takeuchi05,brauer07}.  
Perhaps the most promising option for slowing (or stopping) this transport 
mechanism is with a ``bump" in the radial gas pressure profile 
\citep[e.g.,][]{whipple72}, either locally and stochastically in over-densities 
generated by turbulence \citep[e.g.,][]{klahr97,pinilla12a} or globally and 
with long duration in density concentrations produced near sharp ionization 
boundaries \citep[just outside a ``dead" zone; e.g.,][]{dzyurkevich10} or 
through dynamical interactions with a companion \citep[e.g.,][]{pinilla12b}.  

The most obvious case in which the latter scenario is relevant is for a 
circumbinary disk, where dynamical interactions between the stars and gas 
reservoir clear the disk material inside a radius $\sim$3$\times$ larger than 
the binary separation \citep[e.g.,][]{artymowicz94}.  The steep gas pressure 
gradient created by this clearing will trap particles in a circumbinary 
``ring", which itself might have significantly enhanced pressures due to the 
dynamical excitation of density waves \citep[e.g.,][]{artymowicz96,kley06,
hayasaki09}.  GG Tau is the canonical example of a close young stellar 
pair \citep[$\sim$0\farcs25 projected separation;][]{leinert91,ghez93} with a 
prominent circumbinary ring, which has been resolved and extensively studied in 
mm-wave continuum and molecular line emission \citep{dutrey94,guilloteau99,
pietu11} as well as scattered light in the optical and near-infrared 
\citep{roddier96,silber00,krist02,krist05,mccabe02,itoh02,duchene04}.  The 
(unresolved) radio spectrum indicates that $\sim$mm/cm-sized particles are 
present in the GG Tau circumbinary ring \citep[e.g.,][]{guilloteau99,rodmann06,
scaife13}, lending some additional support to the idea that radial drift is 
halted near the ring edge.  

Here we present resolved measurements of continuum emission from the GG Tau 
system at wavelengths of 1.3, 2.8, 7.3, and 50\,mm, in an effort to 
characterize the dust population in the GG Tau A circumbinary ring.  The 
observations and data calibration are presented in Section 2.  Models of the 
resolved emission and broadband spectrum are developed and analyzed in Section 
3.  The results are discussed in the context of current ideas about the 
evolution of disk solids in Section 4.

\section{Observations and Data Reduction}

GG Tau was observed with the 15-element ($6\times10.4$\,m and $9\times6.1$\,m 
antennas) Combined Array for Research in Millimeter Astronomy (CARMA) in its C 
and B configurations (30--350\,m and 100--1000\,m baselines, respectively) with 
the 1\,mm receivers on 2007 September 17 and November 26, and in the B 
configuration with the 3\,mm receivers on 2008 January 8, January 17, and 
February 15.  In the former, the correlator was set up to process two 500\,MHz 
basebands with coarse spectral resolution in each sideband (2\,GHz of total 
bandwidth), with a local oscillator frequency of 228\,GHz ($\lambda = 
1.31$\,mm).  For the latter, a third 500\,MHz baseband was added to each 
sideband (3\,GHz bandwidth), and the receivers were tuned to 106\,GHz ($\lambda 
= 2.83$\,mm).  The observations alternated between GG Tau and the nearby 
quasars J0530+135, J0510+180, J0431+206, J0449+113, and 3C 111, with a 
source--calibrator cycle time of 12--15\,minutes.  Additional observations of 
Uranus were made for calibration purposes.  The atmospheric conditions were 
generally good, with 230\,GHz opacities of $\sim$0.15--0.20 and 0.2--0.3 during 
the 1.3 and 2.8\,mm observations, respectively.

\begin{figure}[t!]
\epsscale{1.0}
\plotone{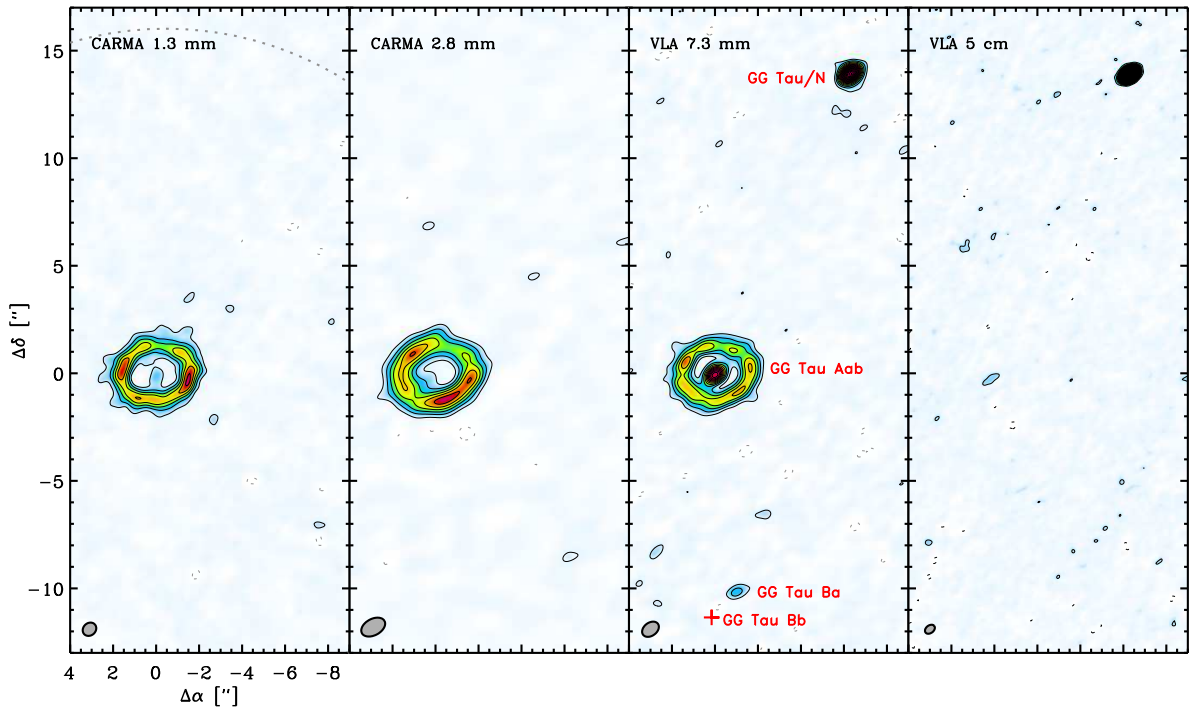}
\figcaption{Synthesized continuum images of the GG Tau field at (from left to
right) wavelengths of 1.3, 2.8, 7.3, and 50\,mm.  Contours are drawn at
3\,$\sigma$ intervals in each panel (7, 1.5, 0.04, and 0.02\,mJy beam$^{-1}$
from left to right), and synthesized beam dimensions are marked in the lower
left corners.  The primary beam for the CARMA 1.3\,mm image is shown as a
dotted gray curve.  The key emission components are labeled in the VLA Q-band
(7.3\,mm) image, including the circumbinary dust ring around GG Tau Aab, faint
radio emission toward the low-mass binary GG Tau Bab (from component Ba), and
the bright radio emission from the background galaxy GG Tau/N.
\label{fig:wide_maps}}
\end{figure}

The raw visibilities were calibrated and subsequently imaged using standard 
tasks in the {\tt MIRIAD} software package.  The passband shape across the 
coarse continuum channels was calibrated using observations of J0530+135, and 
the (antenna-based) complex gain response of the array to both instrumental and 
atmospheric variations was determined from repeated observations of J0530+135 
or 3C 111.  Observations of J0510+180, J0431+206, and J0449+113 were used to 
assess the quality of the gain calibration; the contribution of decoherence due 
to small baseline errors and atmospheric phase noise is found to be small, 
representing a ``seeing" disk with FWHM $\le$0\farcs1.  The absolute amplitude 
scales were set by bootstrapping J0530+135 flux densities from observations of 
Uranus, with systematic uncertainties of $\sim$10\%.  Wideband continuum 
visibilities were created by averaging the spectra across the sampled 
passbands.  These calibrated continuum visibilities were Fourier inverted 
assuming natural weighting, deconvolved with the {\tt CLEAN} algorithm, and 
restored with a synthesized beam to produce the emission maps shown in Figure 
\ref{fig:wide_maps}.  The 1.3\,mm visibilities were tapered with a 0\farcs1 
FWHM Gaussian kernel before inversion to improve the resulting image quality.  
The 1.3\,mm continuum map shown in Figure \ref{fig:wide_maps} has a $0\farcs67 
\times 0\farcs57$ synthesized beam (at P.A.~= 132\degr) and an RMS noise level 
of 2.3\,mJy beam$^{-1}$; the corresponding 2.8\,mm map has a $1\farcs19 \times 
0\farcs76$ beam (at P.A.~= 118\degr) and an RMS noise level of 0.5\,mJy 
beam$^{-1}$.  

New observations of the GG Tau field were also made as part of the 
``Disks@EVLA" key project (project code AC982) with the 27-element (25\,m 
diameter antennas) Karl G.~Jansky Very Large Array (VLA), employing the Q-band 
receivers in the C configuration (35\,m to 3.4\,km baselines) on 2010 November 
27, and the C-band receivers in the A configuration (0.7--36.4\,km baselines) 
on 2011 July 26.  For the Q-band observations, the recently upgraded correlator 
was configured to process two contiguous 1\,GHz basebands centered around 
41.1\,GHz ($\lambda = 7.29$\,mm), each comprising eight 128\,MHz-wide spectral 
windows with 64 channels.  The C-band correlator configuration had a similar 
setup, with the 1\,GHz basebands centered at 4.5 and 7.5\,GHz, for a mean 
frequency of 6\,GHz ($\lambda = 5.0$\,cm).  The observations cycled between GG 
Tau and the nearby calibrator J0431+2037 at $\sim$3 and 10\,minute intervals 
for the Q- and C-bands, respectively.  The bright quasars 3C 84 and 3C 147 were 
also observed for calibration purposes.  

The raw visibilities were calibrated and imaged using the ``Disks@EVLA" 
pipeline in the {\tt CASA} software package (now the standard pipeline for high 
frequency VLA observations\footnote{see \url{https://science.nrao.edu/facilities/vla/data-processing/pipeline}}).  After flagging the data for radio frequency 
interference and other minor issues, the observations of 3C 84 were used to 
calibrate the spectral bandpass response of the system after bootstrapping flux 
densities in each spectral window from observations of 3C 147 (to properly 
treat the shape of the 3C 84 spectrum over the wide relative bandwidth).  The 
Q-band visibilities were then spectrally averaged into 16 pseudo-continuum 
sub-bands (one per 128\,MHz spectral window); the C-band data were not 
averaged, to minimize bandwidth-smearing.  Complex gain variations were 
calibrated with the frequent observations of J0431+2037, and the absolute 
amplitude scale was determined using a frequency-dependent emission model for 
the standard flux density calibrator 3C 147 \citep{perley13}.  The systematic 
uncertainty in the amplitude scale is $\sim$10\%\ at Q-band and 5\%\ at 
C-band.  The calibrated visibilities were Fourier inverted assuming natural 
weighting, deconvolved with the multi-frequency synthesis version of {\tt 
CLEAN}, and restored with a synthesized beam to make the composite continuum 
maps shown in Figure \ref{fig:wide_maps}.  The Q-band map has a 
$0\farcs86\times0\farcs61$ synthesized beam (at P.A.~= 127\degr) and an RMS 
noise level of 13\,$\mu$Jy beam$^{-1}$, and the C-band map has a $0\farcs51 
\times 0\farcs36$ beam (at P.A.~= 127\degr) with an RMS noise level of 
6\,$\mu$Jy beam$^{-1}$.  The C-band image reconstruction required substantial 
extra care, including imaging of the entire primary beam and faceting on 
exceptionally bright background sources, to minimize artifacts at the field 
center.  The images shown in Figure \ref{fig:wide_maps} are not corrected for 
the response of the primary beam.

\section{Analysis and Results} 

\begin{figure}[t!]
\epsscale{1.0}
\plotone{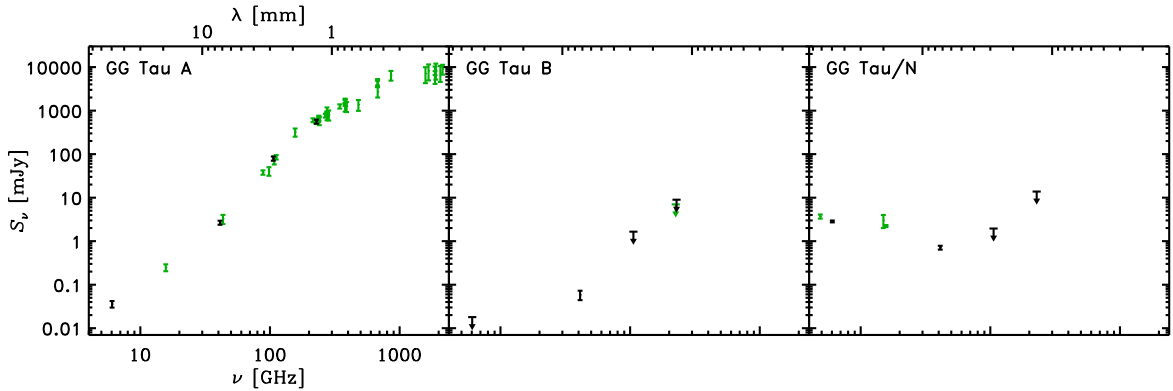}
\figcaption{The mm/radio spectra for the three emission components in the GG 
Tau field: A (left; see Section 3.1 for more details), B (middle), and N 
(right), where the emission from the latter two components were estimated from 
images that have been corrected for the response of the primary beam.  Our 
measurements are shown in black, and flux densities from the literature are 
marked in green (Table \ref{tab:fluxes} lists $S_{\nu}$ for each component).  
The error bars represent the formal statistical uncertainties and the 
systematic calibration uncertainties, added in quadrature.  Upper limits (at 
3\,$\sigma$) are marked with a horizontal line and a downward-pointing arrow. 
\label{fig:spectra}}
\end{figure}

The multifrequency continuum images in Figure \ref{fig:wide_maps} show emission 
from three distinct components: (1) an extended ring around the close binary GG 
Tau A at 1.3, 2.8, and 7.3\,mm, along with point-like emission at its center 
detected at 1.3, 7.3, and 50\,mm;\footnote{It is worth pointing out that we do 
not detect the ``streamer" identified by \citet{pietu11}; its estimated surface 
brightness is comparable to the RMS noise level in our 1.3\,mm map, and would 
likely lie well below the noise floor at longer wavelengths if its origin is 
thermal dust emission.} (2) faint, unresolved emission at 7.3\,mm 
associated with the low-mass star GG Tau Ba (no emission is found toward its 
$\sim$1\farcs5 companion Bb); and (3) bright, unresolved emission at 7.3 and 
50\,mm from the (presumed) extragalactic interloper GG Tau/N.  The composite 
flux densities, $S_{\nu}$, or upper limits for each of these components are 
compiled in Table \ref{tab:fluxes} and displayed together in Figure 
\ref{fig:spectra} (note that the measurements for the GG Tau B and N components 
in this figure and table were determined from images that were corrected for 
the response of the primary beam).  For completeness, we also include 
literature measurements of the radio spectra for each component in this figure.

The radio spectral index of GG Tau/N, $\alpha \approx -0.7$ (defined as 
$S_{\nu} \propto \nu^{\,\alpha}$), and its non-detection at mm wavelengths 
indicate a non-thermal emission mechanism \citep[e.g.,][]{bieging84}.  Given 
this index, GG Tau/N is likely a background extragalactic source (AGN) emitting 
an optically thin synchrotron spectrum; however, it is worth noting that there 
is no confirmed counterpart at any optical or infrared wavelength.  GG Tau B is 
a low-mass T Tauri binary \citep{white99} with a modest infrared excess 
\citep{luhman10}; the 7.3\,mm emission from the primary (Ba; spectral type M5) 
found here is the first detection longward of 24\,$\mu$m.  The nature of this 
Q-band emission from Ba is not clear, given the non-detections at other 
wavelengths.  The derived C-band upper limit allows $\alpha \gtrsim 0.6$, 
consistent with origins in a magnetically active corona \citep{white94,
cranmer13} or dense wind \citep{reynolds86}.  Likewise, non-detections at 1.3 
and 2.8\,mm suggest that $\alpha \lesssim 2.3$, which is also commensurate with 
thermal emission from a relatively cold disk \citep[as might be expected around 
such a low-mass host star; e.g.,][]{andrews13} \hili{or optically thin emission 
from large dust particles \citep[e.g.,][]{ricci10a}}.  Variable non-thermal 
radio emission is an additional (and not mutually exclusive) possibility.

We focus here on the multifrequency, resolved emission structure from GG Tau A, 
itself composed of a narrow circumbinary ring and a point-like contribution 
associated with one or both components of the close binary \citep{dutrey94,
guilloteau99,pietu11}.  In the following sections, we describe a simple model 
to quantify these emission components as a function of the observing frequency, 
and use those results to assess the emission origins.  First, we establish the 
basic spatial structure of the GG Tau A emission (in Section 3.1), and then we 
employ that resolved information to model the spectrum and extract physical 
constraints on the properties of the constituent solid particles (e.g., mass, 
temperature, size distribution; in Section 3.2).

\subsection{Models for Resolved Emission from GG Tau A}

We adopt a simple model prescription for the brightness distribution of the 
GG Tau A emission consisting of an azimuthally-symmetric Gaussian ring and a 
central point source.  The ring is described by seven parameters: a mean radius 
$\mu_r$, width $\sigma_r$, integrated flux density $S_{\nu,r}$ ($= \int 
I_{\nu,r} \, d\Omega$), inclination angle $i$ (0\degr\ is face-on), minor axis 
position angle $\varphi$ (representing the sky-projected orientation of the 
ring rotation axis), and two nuisance parameters to account for the ring center 
location \{$\Delta \alpha_r$, $\Delta \delta_r$\} (defined as arcsecond offsets 
in right ascension and declination from the observed phase center).  The point 
source component includes three additional parameters: a flux density 
$S_{\nu,c}$ and projected offsets \{$\Delta \alpha_c$, $\Delta \delta_c$\} {\it 
relative to the ring center}.  We assume a distance of 140\,pc 
\citep[e.g.,][]{torres09}, and compute models for different frequencies 
independently.

For a given set of these 10 parameters, we calculate synthetic complex 
visibilities, $V_{\nu}$, sampled at the same spatial frequencies, $(u, v)$, as 
the relevant observations.  We then evaluate a Cauchy log-likelihood function 
\citep[cf.,][]{sivia06} to represent the probability of the model, \{$V^{\rm 
M}_{\nu}(u,v)$\}, given the observed complex visibilities, \{$V^{\rm 
D}_{\nu}(u,v)$\}, and their uncertainties, \{$\sigma^{\rm D}_{\nu}(u,v)$\}, 
\begin{equation}
\mathcal{L} \propto \sum_{k} \ln \left( \frac{1-e^{-R_k^2/2}}{R_k^2} \right);   \,\,\,\, {\rm where} \,\,\,\, R_k = \left( \frac{V^{\rm D}_{\nu}(u_k,v_k) - V^{\rm M}_{\nu}(u_k,v_k)}{\sigma^{\rm D}_{\nu}(u_k,v_k)} \right).
\end{equation}
This log-likelihood function was preferred over its more familiar Gaussian 
counterpart (where $\mathcal{L} \propto - \chi^2/2 = \sum_k R_k^2/2$) because 
it is more forgiving of outliers (mitigating parameter bias due to phase 
calibration systematics on long baselines) and more conservative (which is 
important, since our error estimates are derived solely from the visibility 
weights, and therefore do not treat scatter due to issues like pointing 
errors, atmospheric phase noise, etc.).  The posterior probability distribution 
function (PDF) was sampled with Monte Carlo Markov Chain (MCMC) calculations, 
using the \citet{goodman10} ensemble sampler as implemented by 
\citet{foreman-mackey13}.  

A set of initial MCMC calculations was conducted with uniform priors on all 
parameters.  However, given the very weak emission from the central point 
source component at 1.3 and 2.8\,mm, convergence on the relative offset 
parameters \{$\Delta \alpha_c$, $\Delta \delta_c$\} was prohibitively slow (and 
could therefore lead us to biased inferences of $S_{\nu,c}$).  In these initial 
calculations, we found that the 1.3\,mm relative offsets were entirely 
consistent with the well-constrained values at 7.3\,mm.  Therefore, new MCMC 
calculations were made with (independent) Gaussian priors on these offsets, 
centered on the 7.3\,mm values and with standard deviations corresponding to 
the inferred 68\%\ marginal widths of their posterior PDFs.  In any case, this 
had no impact on the inferences for other parameters.

\begin{figure}[t!]
\epsscale{0.93}
\plotone{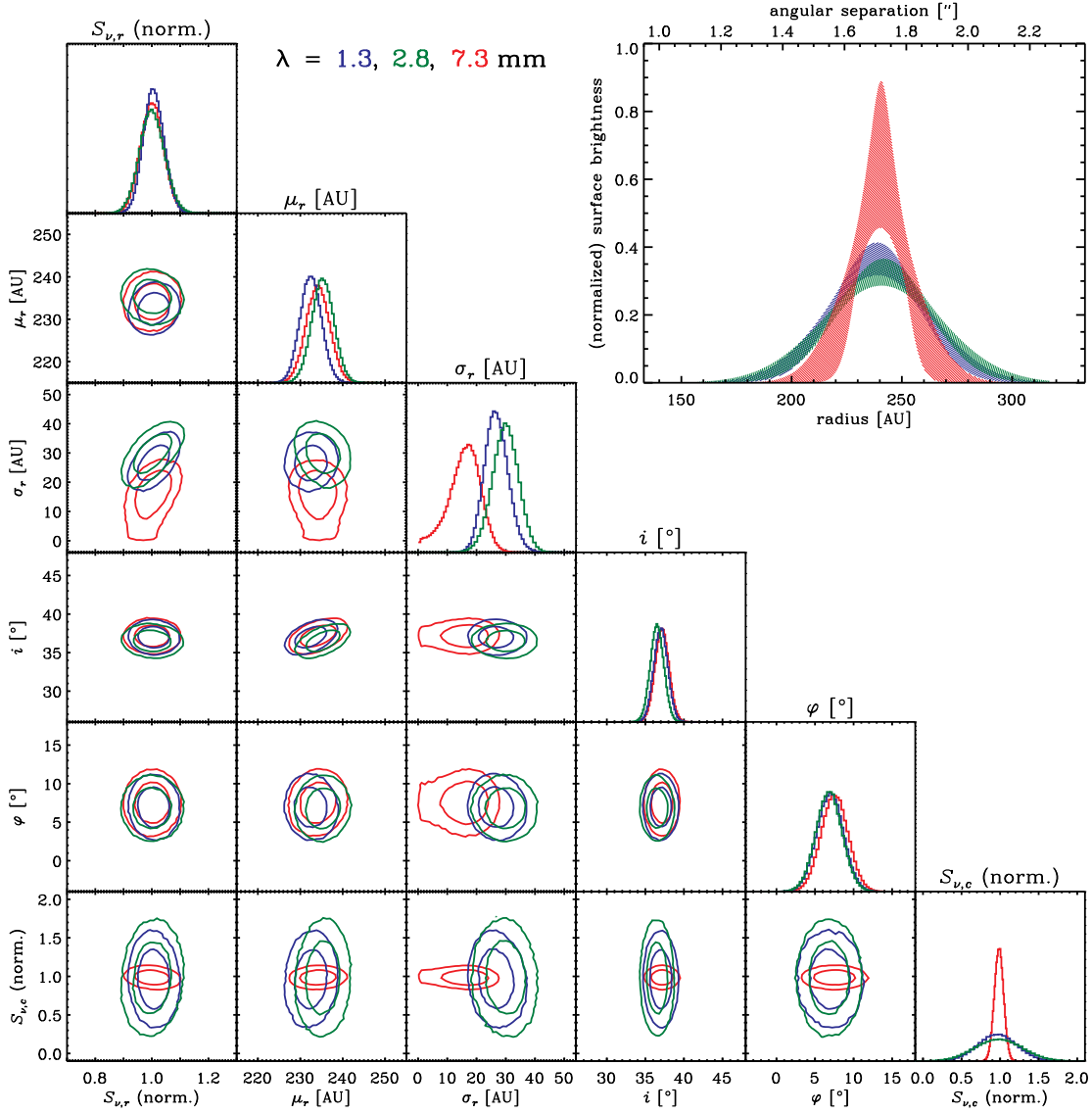}
\figcaption{Summary of the posterior PDFs inferred from MCMC calculations in 
reference to resolved interferometer data at 1.3, 2.8, and 7.3\,mm, assuming a 
model composed of a central point source and a Gaussian ring.  The staircase 
plot to the left shows marginalized two-parameter posterior PDF surfaces, with 
contours drawn at 68 and 95\%\ confidence intervals.  Marginalized posterior 
PDFs for each parameter are shown along the diagonal.  The ring and point 
source flux densities, $S_{\nu,r}$ and $S_{\nu,c}$ respectively, are normalized
by their best-fit values for clarity.  Note that the 4 directional offset 
parameters ($\Delta \alpha_r$, $\Delta \delta_r$, $\Delta \alpha_c$, and 
$\Delta \delta_c$) are not shown, to simplify the plot.  The top right panel is 
a visual representation of the (normalized) radial surface brightness profiles 
derived from this analysis; the widths of these profiles represent the 68\%\ 
confidence intervals.  \label{fig:mcmc}}
\end{figure}

\begin{figure}[t!]
\epsscale{0.95}
\plotone{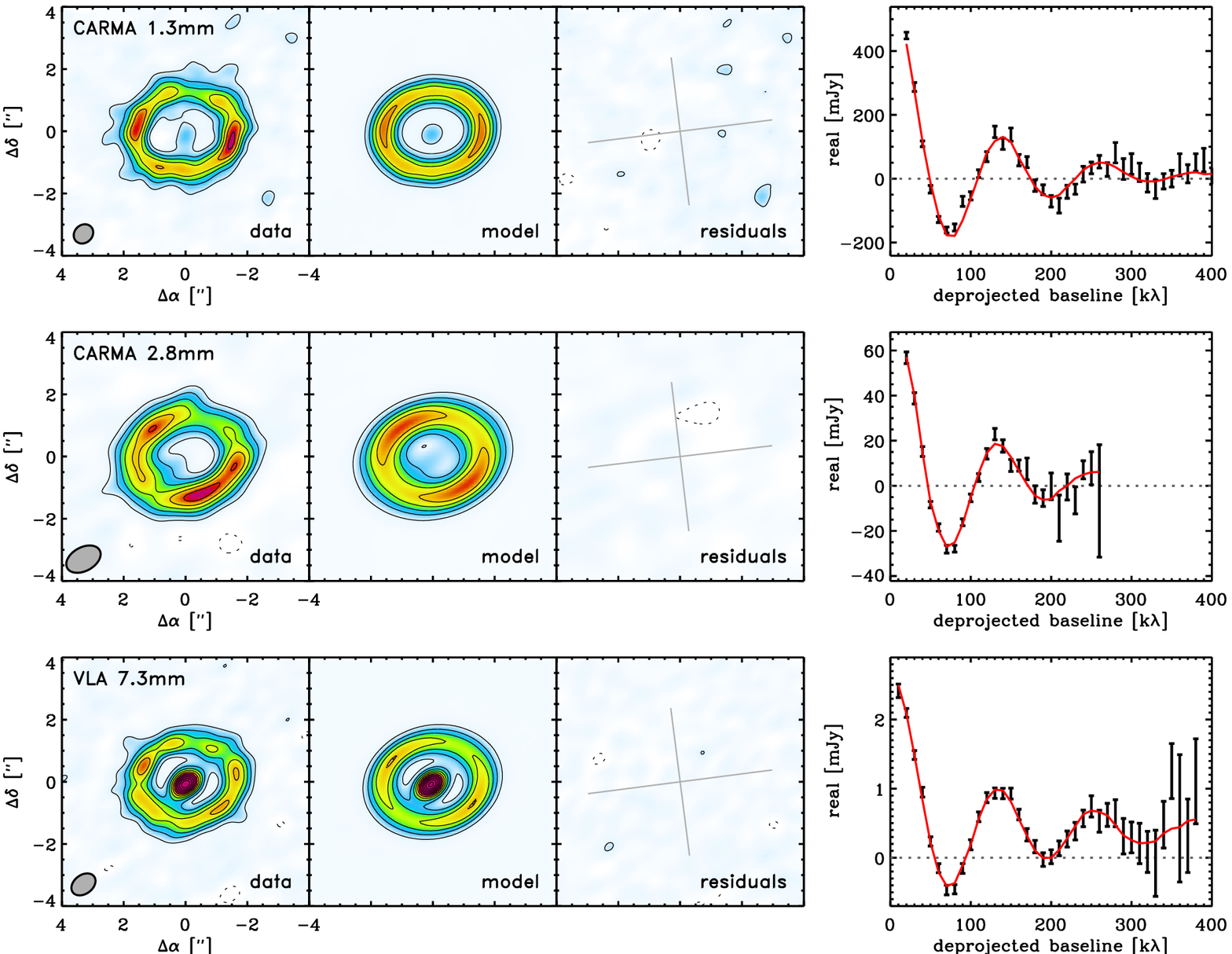}
\figcaption{Comparisons of the data and best-fit models.  The left-hand panels 
show the data and the corresponding images synthesized from the model and 
residual visibilities in the same way as the data.  Contour levels are as in 
Figure \ref{fig:wide_maps}.  A cross marks the ring center and orientation in 
the residuals panel.  The rightmost panel in each row shows the 
azimuthally-averaged (real) visibility profiles, deprojected according to the 
derived ring geometry (data in black, models in red). 
\label{fig:dmr}}
\end{figure}

Figure \ref{fig:mcmc} is a representation of the sampled posterior PDFs in the 
form of a staircase diagram, showing both pair-wise parameter covariances 
(contours are drawn at the 68 and 95\%\ confidence intervals) and marginalized 
posterior PDFs for individual parameters (the four offset parameters are not 
shown, and the flux density parameters are normalized to the peaks of their 
marginal posterior PDFs, for the sake of clarity).  The panel in the upper 
right corner is a visualization of the (area-normalized) radial surface 
brightness profiles reconstructed from random draws to the joint posterior 
PDFs, where the shaded widths of each profile are representative of the 68\%\ 
(i.e., $\sim$1\,$\sigma$) confidence intervals.  Table \ref{tab:model} 
summarizes the modeling results, listing the ``best-fit" (peaks of the marginal 
posterior PDFs) parameter values and their 68\%\ confidence intervals at each 
observing frequency.  A direct comparison of the data and best-fit models is 
shown in Figure \ref{fig:dmr}, in both the image plane and with the 
azimuthally-averaged (and deprojected) visibility profiles.  

Within the uncertainties, the GG Tau A circumbinary ring has the same mean 
radius and width (as well as inclination, orientation, and center) at all 
observed frequencies; $\mu_r \approx 230$--240\,AU ($\sim$1\farcs7) and 
$\sigma_r \approx 20$--30\,AU (corresponding to a FWHM of 0.3--0.5\arcsec; 
i.e., it is only marginally resolved).  There is a hint that the ring is 
slightly narrower (at the $\sim$1\,$\sigma$ level) at 7.3\,mm, although the 
difference is not statistically (or practically) significant.  Taken together, 
this indicates that the spectral behavior of the dust continuum emission {\it 
does not vary} radially across the ring \hili{on the angular scales and at the 
spectral sensitivity currently available.  If we assume that a Gaussian 
distribution is an appropriate spatial model and a power-law with frequency is 
a reasonable spectral model, we can very roughly estimate from the fitting 
results derived here that $\Delta \alpha \lesssim 0.3$ on (radial) angular 
scales $\gtrsim$0\farcs1.  Variations at finer scales are possible, and perhaps 
likely (see Section 4).} 

There is some non-negligible curvature in the ring spectrum, with a steeper 
spectral index at lower frequencies: $\alpha$(41--106\,GHz) $\approx 3.7\pm0.2$ 
and $\alpha$(106--228\,GHz) $\approx 2.6\pm0.2$ (the quoted uncertainties 
account for the $\sim$10\%\ systematics introduced in the amplitude 
calibrations).  The peak brightness temperature of the ring at 1.3\,mm is only 
$\sim$0.2\,K, confirming that the emission is optically thin.  Therefore, the 
observed spectral curvature is a by-product of the intrinsic shape of the dust 
opacity spectrum (and perhaps cool temperatures; see Section 3.2 for more 
details).  The ring is not detected at a wavelength of 5\,cm.  Assuming it has 
the same emission morphology as at shorter wavelengths, a limit on its 
integrated flux density can be made based on the measured RMS noise level 
(6\,$\mu$Jy beam$^{-1}$) in the C-band map: we estimate $S_{\nu,r} < 
40$\,$\mu$Jy (3\,$\sigma$).  

The central point-like component, originally identified at 1.4 and 1.1\,mm by 
\citet{guilloteau99} and \citet{pietu11}, is also detected here at 1.3, 7.3, 
and 50\,mm.  Models with a relatively faint central source at 2.8\,mm (as 
listed in Table \ref{tab:model}) provide better overall matches to the 
visibility data, although formally the inference on $S_{\nu,c}$ is only 
marginally significant (greater than zero at the $\sim$2.7\,$\sigma$ level): 
alternatively, we could quote a 3\,$\sigma$ upper limit as $S_{\nu,c} < 
4.0$\,mJy.  The (sparse) radio spectrum of this central source includes 
contributions from different emission mechanisms.  The spectral index at high 
frequencies is steep, $\alpha$(41-228\,GHz) $\approx 2.1\pm0.2$, and consistent 
with optically thick thermal emission from a warm dust disk.  The low-frequency 
radio spectrum is considerably more shallow, $\alpha$(6--41\,GHz) $\approx 
1.3\pm0.1$; when combined with the thermal spectrum, the radio emission is best 
explained with an intrinsically flat spectrum ($\alpha \approx 0$), suggesting 
a contribution from optically thin free-free radiation.  We find no evidence 
that this central component is resolved: models of the 7.3\,mm emission that 
assume a Gaussian emission distribution (rather than a point source) indicate a 
radial width $<$12\,AU (3\,$\sigma$; this corresponds to a HWHM $< 0\farcs2$).  
However, our models suggest that this component is marginally offset from the 
ring center, with a sky-projected separation of $90\pm30$\,mas to the southeast 
(at P.A. $\approx 149\pm1$\degr) measured from the 7.3\,mm data.  Assuming the 
recent orbit calculations of \citet{koehler11}, and associating the ring center 
with the projected binary center of mass \citep[with the $\sim$0.9:1 mass ratio 
of][]{white99}, these constraints on the scale and orientation of the offsets 
suggest that the emission likely originates from the primary component GG Tau 
Aa.  A similar inference was made for the 1.1\,mm emission detected by 
\citet{pietu11}, based on the orbit calculations of \citet{beust05}.

\subsection{Models of the GG Tau A Spectrum}

Having established an empirical model of the resolved multifrequency emission 
from GG Tau A, we shift focus to develop a more physically-motivated model of 
the entire far-infrared to radio spectrum.  The basic goal is to help 
characterize the dust population in the circumbinary ring.  The emission 
structure of GG Tau A that we derived in the previous section is unusually 
simple in the context of protoplanetary disks.  The circumbinary ring has a 
well-defined mean radius and width, and is spatially isolated from the central 
component.  The absence of a spatial gradient in the ring spectrum, along with 
its apparently low optical depth, suggest that the continuum emission is a 
reasonably good tracer of the dust column density distribution.  Moreover, the 
ring itself is sufficiently narrow and distant from the central heating sources 
that we expect it should have a near-constant radial temperature profile 
(irradiation heating would produce a variation of $\lesssim$\,15\%, roughly 
2--3\,K, across the ring; see below).  So unlike a typical disk, where large 
and uncertain gradients in temperature and density can act as severe obstacles, 
we have an interesting opportunity to use the mm/radio spectrum and our 
structural constraints from the resolved data to extract some constraints on 
the size distribution of the solids in the circumbinary ring.

To that end, we define a spectrum model as the sum of two components: thermal 
dust emission in the ring (with flux densities $S_{\nu,r}$) and a composite 
emission origin associated with the central point source ($S_{\nu,c}$).  For 
the latter, we assume a double power-law spectrum,
\begin{equation}
S^{\rm M}_{\nu,c} = S_{\nu,0}^{\rm dust} \left(\frac{\nu}{{\rm 10\,GHz}}\right)^{\alpha_{\rm dust}} + S_{\nu,0}^{\rm ff} \left(\frac{\nu}{{\rm 10\,GHz}}\right)^{\alpha_{\rm ff}}.  
\end{equation}
And for the dust emission in the ring, we use the classic and simple 
one-dimensional thermal continuum model often adopted for disks \citep{adams87,
beckwith90}, 
\begin{equation}
S^{\rm M}_{\nu,r} = \frac{2\pi\cos{i}}{d^2} \int dr\,r\,B_{\nu}(T_r)\,(1 - e^{-\kappa_{\nu}\Sigma_r/\cos{i}}), 
\end{equation}
where $d = 140$\,pc, $i$ is the ring inclination, $B_{\nu}(T_r)$ is the Planck 
function at a given temperature, $\Sigma_r$ is a Gaussian surface density 
profile with mean $\mu_r$, width $\sigma_r$, and peak value $\Sigma_0$, and 
$\kappa_{\nu}$ is the dust opacity spectrum.  Model opacity spectra were 
derived assuming a power-law grain size ($a$) distribution, with index $q$ 
(where $dn/da \propto a^{-q}$) and maximum size $a_{\rm max}$ (the minimum size 
was set to 0.1\,$\mu$m).  For easier comparisons with related work, we 
employed the \citet{ricci10a} dust mixture\footnote{From a material composition 
standpoint, this mixture is similar to the one advocated by 
\citet{pollack94}.}, with volume fractions of 30\%\ vacuum (porosity), 7\%\ 
silicates, 21\%\ carbonaceous materials, and 42\%\ water ice, using optical 
constants from \citet{weingartner01}, \citet{zubko96}, and \citet{warren84}, 
respectively.  The optical constants for the mixture were determined with the 
Bruggeman rule, and the corresponding $\kappa_{\nu}$ for any particle size were 
computed with a Mie code.  Altogether, the model has 11 parameters, 
\{$S_{\nu,0}^{\rm dust}$, $\alpha_{\rm dust}$, $S_{\nu,0}^{\rm ff}$, 
$\alpha_{\rm ff}$, $i$, $T_r$, $\Sigma_0$, $\mu_r$, $\sigma_r$, $a_{\rm max}$, 
$q$\}.  

For any parameter combination, we define two residual terms at each frequency 
that record the fit quality relative to the {\it resolved} measurements of the 
ring and point source flux densities, 
\begin{equation}
R_{\nu,r} = \left(\frac{S^{\rm D}_{\nu,r}-S^{\rm M}_{\nu,r}}{\sigma^{\rm D}_{\nu,r}}\right) \,\, {\rm and} \,\,\,\, R_{\nu,c} = \left(\frac{S^{\rm D}_{\nu,c}-S^{\rm M}_{\nu,c}}{\sigma^{\rm D}_{\nu,c}}\right),
\end{equation}
respectively; the ``observed" flux densities, $S^{\rm D}_{\nu}$, and their 
associated uncertainties, $\sigma^{\rm D}_{\nu}$, can be found in Table 
\ref{tab:model}.\footnote{Here we also add in quadrature a systematic 
calibration uncertainty term at each frequency; see Section 2.}  Moreover, we 
made use of the {\it unresolved} photometry in the literature to compare with 
the sum of the model components, with an additional residual term
\begin{equation}
R_{\nu,{\rm tot}} = \left(\frac{S^{\rm D}_{\nu,{\rm tot}}-[S^{\rm M}_{\nu,r}+S^{\rm M}_{\nu,c}]}{\sigma^{\rm D}_{\nu,{\rm tot}}}\right),
\end{equation}
at each frequency, where \{$S^{\rm D}_{\nu,{\rm tot}}$, $\sigma^{\rm 
D}_{\nu,{\rm tot}}$\} are compiled in Table \ref{tab:fluxes}.  These terms are 
treated as a combined residual, $R_{\nu} = \{R_{\nu,r}, R_{\nu,c}, R_{\nu,{\rm 
tot}}\}$, in evaluating a log-likelihood function as in Eq.~(1) (now the 
summation is over $\nu$).  The same MCMC algorithm utilized in Section 3.1 was 
employed to optimize the model and sample the posterior PDFs for the model 
parameters.  Because there are more parameters than constraints describing the 
point source, we adopted Gaussian priors on \{$\log{S_{\nu,0}^{\rm dust}}$, 
$\alpha_{\rm dust}$, $\log{S_{\nu,0}^{\rm ff}}$, $\alpha_{\rm ff}$\}, with 
means \{-4.65, 2.1, -4.25, 0.0\} and widths \{0.2, 0.2, 0.2, 0.1\} (the 
normalizations are in log\,Jy units) informed by the examination of the 
$S_{\nu,c}$ spectrum described above.  To incorporate the measurements of the 
resolved emission structure found in Section 3.1, we assumed Gaussian priors 
for \{$\mu_r$, $\sigma_r$, $i$\} with means \{235\,AU, 25\,AU, 37\degr\} and 
widths \{5\,AU, 5\,AU, 1\degr\} (see Table \ref{tab:model}).  Uniform priors 
were assumed for the other parameters, \{$T_r$, $\Sigma_0$, $a_{\rm max}$, 
$q$\}.

\begin{figure}[t!]
\epsscale{0.7}
\plotone{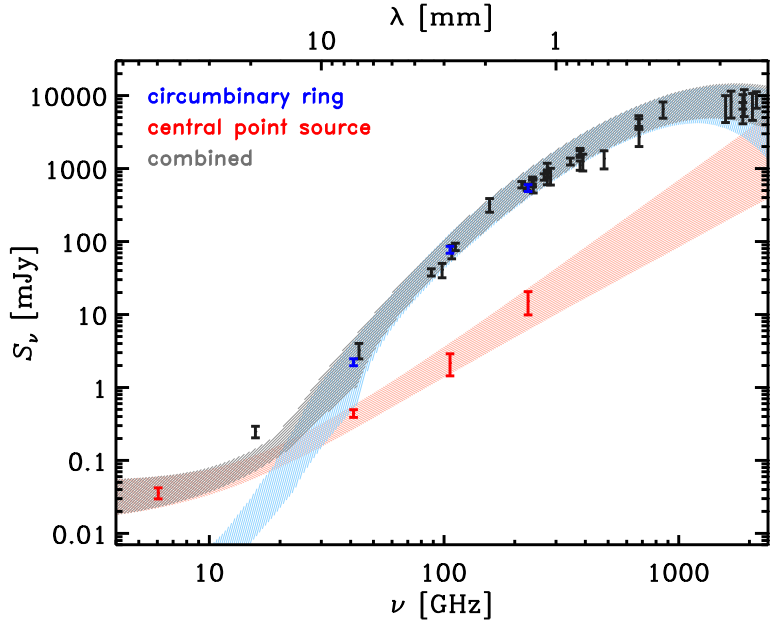}
\figcaption{The far-infrared to radio spectrum of GG Tau A, decomposed into the
circumbinary ring ({\it blue}) and central point source ({\it red}), along with
their combination ({\it gray}).  The model spectrum contributions from each 
component, based on random draws from the posterior PDF, are overlaid on the 
data; their widths represent the 95\%\ ($\sim$2\,$\sigma$) confidence 
boundaries.  \label{fig:modelSED}}
\end{figure}

\begin{figure}[t!]
\epsscale{0.7}
\plotone{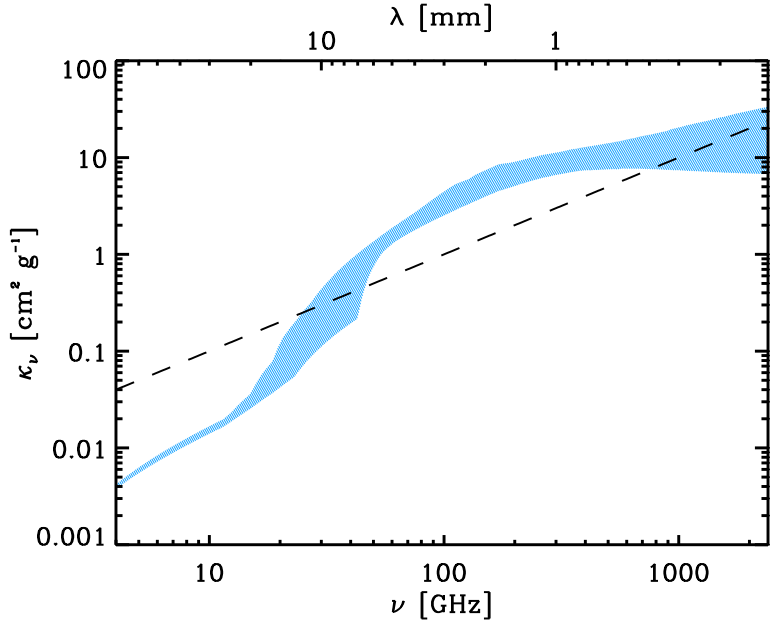}
\figcaption{Constraints on the dust opacity spectrum ($\sim$2\,$\sigma$ 
confidence intervals) in the GG Tau A circumbinary ring, reconstructed from 
random draws of the joint posterior PDF derived from modeling the far-infrared 
to radio spectrum.  Overlaid as a dashed curve is the standard opacity 
prescription used for protoplanetary disks, originally advocated by 
\citet{beckwith90}.  \label{fig:opacities}}
\end{figure}

The parameter inferences from these fits are summarized in Table 
\ref{tab:spec}.  The corresponding model spectra are compared with the data in 
Figure \ref{fig:modelSED}.  In terms of the physical conditions in the ring, we 
infer a dust temperature of 20--30\,K, comparable to the 35\,K determined from 
the CO spectral line emission by \citet{guilloteau99}.  The difference is 
plausibly a manifestation of the modest temperature inversion that would be 
expected between the cooler midplane (where the dust emission is generated) and 
the warmer atmosphere traced by the CO \citep[e.g.,][]{dartois03,
rosenfeld13a}.  The dust surface density at the peak of the ring is only 
0.01--0.03\,g\,cm$^{-2}$, implying a total dust mass of 
$\sim$30--90\,M$_{\oplus}$ (or 1--3$\times10^{-4}$\,M$_{\odot}$).  We find that 
relatively top-heavy grain size distributions, $q \approx 1.4$--2.7, provide 
substantially better fits than the typical assumptions based on models of 
collisional cascades \citep{dohnanyi69} or measurements in the diffuse 
interstellar medium \citep{mathis77}, where $q \approx 3.5$.  With more mass 
concentrated near the maximum particle size, $a_{\rm max} \approx 1$--2\,mm, 
the corresponding dust opacity spectrum ends up having substantial curvature at 
wavelengths near $a_{\rm max}$, reproducing well the observed shape of the 
mm-wave ring spectrum.  A reconstruction of the inferred $\kappa_{\nu}$ is 
shown in Figure \ref{fig:opacities}: the 1.3\,mm opacity lies in the range of 
5--10\,cm$^2$\,g$^{-1}$, $\sim$2--4$\times$ larger than the standard 
\citet{beckwith90} opacity prescription for disks.  Note that the ring is 
optically thin at all frequencies of interest here.

The best-fit model has a $\chi^2 \approx 62$, and a reduced $\tilde{\chi}^2 
\approx 1.8$ (46 datapoints, 11 free parameters).\footnote{Note that we do not 
double-count the ring+central source photometry measured here in the 
``combined" model fit, despite listing those measurements in Table 
\ref{tab:fluxes} (for the sake of completeness).}  Some of these residuals are 
likely due to under-estimates of flux density uncertainties, particularly for 
older single-dish photometers at challenging submillimeter wavelengths.  It is 
worth explicitly pointing out that the favored models systematically 
under-predict the Ku-band (16\,GHz) flux density reported by \citet{scaife13} 
at the $\sim$3\,$\sigma$ level (this datapoint alone drives $\chi^2$ up by 
$\sim$10).  Perhaps this is due to the difficulty of disentangling the GG Tau A 
and N emission in those data,\footnote{\hili{As well as any (relatively) small 
contribution from GG Tau B.}} where the resolution was nearly 3$\times$ the 
A--N separation and the contrast ratio is high (in her Figure 1, Scaife 
indicates that N is roughly an order of magnitude brighter than A at this 
frequency).  Alternatively, it is possible that the measurement uncertainties 
are fine, and instead some of the assumptions made in the modeling are 
responsible for the inferred (modest) residuals.  To explore that possibility 
further, we re-fit the data with several modifications to our critical 
assumptions.  

In motivating the simplicity of the GG Tau A circumbinary ring structure, we 
suggested that its dust temperature is roughly constant.  However, a radial 
temperature gradient could be present, particularly if the inner part of the 
ring intercepts a substantial amount of incident irradiation from the central 
stars \citep[analogous to the dust sublimation boundaries of most disks, or the 
``wall" features just outside transition disk cavities; e.g.,][]{dullemond01,
calvet02}.  As an extreme counter-example, we re-modeled the spectrum assuming 
that the dust temperatures vary inversely with radius in the ring ($T_r \propto 
1/r$).  No significant differences in the model parameters were found, as might 
be expected given the narrowness of the ring.  This check validates the 
simplified assumption that reasonable temperature gradients have negligible 
impact on our results.

The model optimization described above currently requires relatively stringent 
priors on the central point source parameters (given the sparse data).  But 
there is considerable leeway in those prior assumptions that could still offer 
reasonable fits to the data.  To assess the impact of these priors on the main 
physical parameters of the circumbinary ring, we re-modeled the spectrum with a 
steeper combination of Gaussian priors on the spectral indices, with means 
\{$\alpha_{\rm dust}$, $\alpha_{\rm ff}$\} = \{2.5, 1.0\}, and adjustments to 
their corresponding normalizations (at 10\,GHz), \{$\log{S^{\rm 
dust}_{\nu,0}}$, $\log{S^{\rm ff}_{\nu,0}}$\} = \{-5.25, -4.25\} (in Jy units; 
the widths of the priors were the same as used above).  We find that adopting 
these alternative priors has no notable effect on the key ring parameters; the 
point source is simply too faint to matter in this regard (nor does it improve 
the fit quality, even for the Ku-band point noted above).  That said, the 
uncertain origins of the central point source emission will remain an open 
question until resolved measurements at additional frequencies are available.

Perhaps the more relevant assumptions made in the modeling concern the nature 
of the grains themselves, particularly their porosities and material 
compositions.  To investigate these issues further, we first re-modeled the GG 
Tau A spectrum assuming different (fixed) porosities.  Compared to our nominal 
parameters (30\%\ porosity), models with ``compact" grains (0\%\ porosity) can 
explain the data reasonably well if the circumbinary ring has a similar $T_r$, 
a slightly ($\sim$30--50\%) lower dust mass, and a steeper size distribution 
($q \approx 2.5$--3.7) with $a_{\rm max}$ just under 1\,mm.  For higher 
porosities (60\%), we instead infer a (5--8$\times$) higher dust mass and 
shallow size distribution ($q \approx 1.0$--2.8) up to a maximum size of a few 
mm to $\sim$1\,cm.  Since $a_{\rm max}$ is comparable to the wavelengths of 
interest, resonances preferentially enhance the mm-wave opacities for grains 
with higher filling factors \citep[lower porosities; e.g.,][but see 
\citealt{cuzzi14}]{miyake93,kataoka14}: a higher $\kappa_{\nu}$ drives us to 
find lower $\Sigma_0$ (and therefore dust mass) for a fixed spectrum (and vice 
versa).  These same resonances also naturally introduce an intrinsic curvature 
into the mm/cm-wave opacity spectrum that is reflected in the inferred size 
distribution index ($q$): since compact grains then do not require the added 
$\kappa_{\nu}$ curvature produced by a top-heavy size distribution, it makes 
sense that we infer a higher $q$ when the porosity is low (and vice versa).  
For a fixed composition, grains with $\sim$20--40\%\ porosity do produce better
fits to the data.  

Next, we performed a similar experiment that varied the material composition of 
the grains, by re-fitting the data with extreme silicate- or carbon-rich 
mixtures.\footnote{Since water ice plays only a minor role relative to 
silicates and carbonaceous material in setting $\kappa_{\nu}$ in the mm/cm 
wavelength range, its volume fraction is left fixed \citep[at 42\%; 
cf.,][]{pollack94} in this experiment.}  Compared to our adopted mixture (with 
7\%\ silicates and 21\%\ carbonaceous material by volume), models with C-rich 
grains (28\%\ carbon, no silicates) in the GG Tau A ring have a comparable dust 
mass and temperature, and a modestly steeper grain size distribution ($q 
\approx 2.2$--3.5) up to $a_{\rm max} \approx 1$\,mm.  In the opposite case, 
models with Si-rich grains tend toward ($\sim$10$\times$) higher dust masses 
and shallow size distributions ($q \approx 1.1$--2.6) with $a_{\rm max} \approx 
2$--8\,mm.  Silicate-rich grains have mm/cm-wave opacity spectra that exhibit 
less curvature and are preferentially reduced compared to C-rich grains, 
properties that naturally account for our inferences of top-heavy size 
distributions and higher dust masses when C is depleted, respectively.  That 
said, models relying on these extreme compositions produce poor fits to the 
data; the experiment is only intended to convey the {\it sense} of the 
variation.  

While there are many alternative assumptions along these lines that could be 
made in the modeling, the investigations described above demonstrate that most 
of the key parameters related to the dust in the GG Tau A circumbinary ring are 
relatively robust.  Within a reasonably large range of temperature gradients, 
point source emission models, grain porosities and compositional variations, we 
find dust temperatures of $\sim$20--30\,K, dust masses of 
$\sim$1--10$\times$10$^{-4}$\,$M_{\odot}$, and maximum grain sizes of 
$\sim$1--10\,mm (in all cases with preference near the lower end of the quoted 
ranges).  However, it should be clear that these data do not provide a strong 
quantitative constraint on the power-law index of the grain size distribution, 
$q$.  Finally, as a point of reference, we find that the dust opacity at 
1.3\,mm is restricted to $\sim$0.2--20\,cm$^2$\,g$^{-1}$ in all the models 
explored here.

\section{Discussion and Conclusions}

We have obtained and analyzed high angular resolution observations of continuum 
emission associated with GG Tau, a young quadruple star system, at wavelengths 
of 1.3, 2.8, 7.3, and 50\,mm.  These data confirm that GG Tau/N is a background 
synchrotron source, and identify faint 7.3\,mm emission associated with the 
low-mass star GG Tau Ba, although the origin of the latter is unclear.  As had 
been noted previously \citep[e.g.,][]{guilloteau99}, we find a bright emission 
ring and central peak associated with GG Tau A (although the ring is undetected 
at a wavelength of 5\,cm).  The visibility data were modeled with a simple 
surface brightness prescription.  We found that the emission at all frequencies 
could be described well using a (slightly offset) central point source and a 
(projected) circular Gaussian ring with a mean radius ($\mu_r$) of 
$235\pm5$\,AU and width ($\sigma_r$) of $25\pm5$\,AU (a FWHM of $60\pm10$\,AU), 
reasonably consistent with previous constraints based on slightly different 
assumptions \citep{dutrey94,guilloteau99,pietu11}.  These morphological 
constraints indicate that there is negligible radial variation \hili{(on 
angular scales $\gtrsim0\farcs1$)} of the mm/cm-wavelength spectrum 
\hili{($\alpha \lesssim 0.3$, from 1.3--7.3\,mm)} across the circumbinary ring.

The spectrum of the point source flattens considerably at the longest radio 
wavelengths, suggesting emission contributions from both dust and free-free 
radiation.  The integrated spectrum of the ring structure exhibits 
substantial curvature, becoming steeper at longer wavelengths, which we suggest 
is a manifestation of the intrinsic shape of the dust opacity spectrum.  We 
developed some simple physical models of the observed GG Tau A spectrum, and 
found reasonably good fits for dust temperatures ($T_r$) of 20--30\,K, dust 
masses of 30--60\,$M_{\oplus}$ (1--3$\times10^{-4}$\,$M_{\odot}$), and 
relatively top-heavy grain size distributions ($dn/da \propto a^{-q}$, with $q 
\approx 1.4$--2.7) up to maximum particle sizes ($a_{\rm max}$) of 
$\sim$1--2\,mm, for the grain composition of \citet{ricci10a} (effectively the 
\citealt{pollack94} mixture with 30\%\ porosity).  Alternative assumptions 
about the mineralogical makeup and porosities of the grains permit a much wider 
range of the size index $q$, but still suggest similar temperatures and a 
relatively narrow range of masses ($\sim$1--10$\times10^{-4}$\,$M_{\odot}$) and 
maximum particle sizes ($\sim$1--10\,mm).  In any case, the opacity spectrum 
inferred in the ring is not described well with a single power-law in the mm/cm 
wavelength range; the standard assumption that $\kappa_{\nu} \propto 
\nu^{\beta}$ is not appropriate here.  Although we have estimated significantly 
lower masses \citep[up to a factor of $\sim$5 compared to, 
e.g.,][]{guilloteau99,aw05} and maximum grain sizes \citep[$\sim$4--40$\times$ 
smaller than inferred by][]{scaife13} in the ring than in previous studies, 
these properties still seem remarkably high given the large separation from the 
central stars, narrow radial width of the ring, and age of the system 
($\sim$1--3\,Myr; or rather time available for particle growth).  

\begin{figure}[t!]
\epsscale{0.7}
\plotone{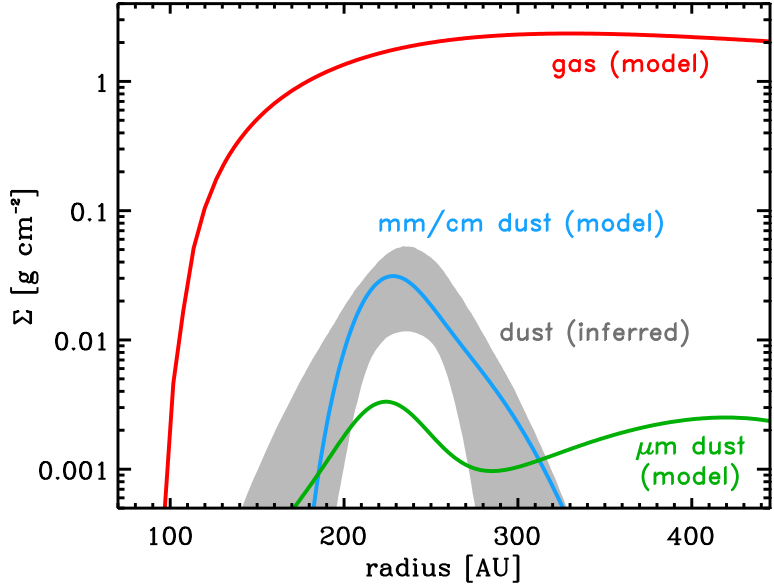}
\figcaption{A schematic snapshot for an illustrative model of dust evolution in 
a truncated gas disk with a gradually tapered inner edge, tuned to the relevant 
parameters of the GG Tau A circumbinary disk.  We assumed a static power-law 
gas surface density profile ($\Sigma_{\rm gas} \propto 1/r$; {\it red} curve), 
with an ``edge" near the circumbinary disk truncation radius expected from 
estimates of the GG Tau A orbit, and a gradual Gaussian taper (with mean 
100\,AU and width 120\,AU).  The total gas mass was 0.13\,$M_{\odot}$ 
($\sim$10\%\ of the binary mass), consistent with the estimate of 
\citet{guilloteau99}.  A population of small (0.1--1\,$\mu$m, with $dn/da 
\propto a^{-3.5}$) grains with radially constant dust-to-gas mass ratio (0.01) 
was evolved in size and space for 1\,Myr, following the \citet{birnstiel10} 
prescription for a constant turbulent viscosity parameter ($\alpha = 0.002$) 
and fragmentation velocity (10\,m s$^{-1}$).  The surface density distributions
of $\mu$m-sized (0.5--5\,$\mu$m; {\it green}) and mm/cm-sized (500\,$\mu$m 
$<a<$ 5\,cm; {\it blue}) particles predicted by this model are in reasonably 
good agreement with both infrared scattered light observations 
\citep[e.g.,][]{duchene04} and our estimates of $\Sigma_{\rm dust}$ inferred 
from the modeling in Section 3 (the {\it gray} shaded region corresponds to the 
2\,$\sigma$ uncertainties).  \label{fig:bump}}
\end{figure}

A promising way to concentrate and grow dust particles up to $\sim$mm sizes at 
such large distances, as well as to halt their normally fast inward migration 
due to radial drift, relies on creating a ``trap" in a local enhancement of the 
gas pressure profile \citep[cf.,][]{whipple72}.  For GG Tau A, this local 
pressure maximum would likely be induced by dynamical interactions between the 
binary and disk edge \citep[e.g.,][]{artymowicz94}.  But models of the GG Tau A 
stellar orbit highlight a potential issue with this interpretation, since the 
ring edge predicted by dynamical simulations lies well inside (at a radius of 
roughly $\sim$100\,AU) the edge location inferred from dust observations 
\citep[e.g.,][]{mccabe02,beust05,koehler11}.  \citet{pinilla12b} have suggested 
that a dust trap could reside substantially beyond the nominal disk ``edge" if 
the gas pressures decrease (relatively) gradually toward the inner part of the 
ring.  To schematically illustrate this point in the case of GG Tau, we used 
the treatment of \citet{birnstiel10} to simulate the evolution of dust 
particles \hili{--- both spatially and in size, subject to growth, 
fragmentation, viscous diffusion, and radial drift ---} in a gas disk that has 
a (static) power-law surface density profile with a Gaussian tapered edge at a 
radius of 100\,AU.  Tuning the shape of this taper (i.e., the width of the 
Gaussian) can push the gas pressure maximum back to a radius comparable to the 
observed dust ring location.  For a reasonable turbulent viscosity parameter 
($\alpha \approx 0.002$) and density normalization (a total gas mass of 
$\sim$0.1\,$M_{\odot}$), we can also reproduce the inferred maximum particle 
size and width of the dust ring on appropriate timescales ($\sim$1\,Myr).  
\hili{This simulation is also in qualitative agreement with the infrared 
scattered light geometry of the GG Tau A ring \citep[e.g.,][]{duchene04}, in 
that it predicts a similar distribution of $\mu$m-sized grains produced via 
fragmentation in the pressure maximum.}  Figure \ref{fig:bump} summarizes this 
demonstrative \hili{(although by no means unique)} example.  

Of course, it is not clear if such ``soft" edges are physically realistic for 
gas-rich circumbinary disks like the one around GG Tau A; detailed hydrodynamic 
simulations \citep[e.g.,][]{bate00,gunther02,kley08} tuned to this specific 
case would be required to assess the feasibility and internal consistency of 
the basic scenario proposed above \citep[although see][]{beust05}.  Regardless, 
the key point is that we should be actively considering the coupled evolution 
of gas and solids in this and similar disks when attempting to reconcile models 
of their tidal truncation with the (still uncertain) stellar orbital 
configurations.  In principle, one could test the proposed explanation with 
sensitive observations of an optically thin line tracer that tracks the gas 
densities near and interior to the dust disk edge at $\sim$0\farcs1 
resolution.\footnote{It is worth noting that this would also require a proper 
accounting of the gas temperatures, which might not be trivial in such a 
relatively dust-poor environment \citep[see][]{bruderer13}.}  In that context, 
it is compelling to note that \citet{guilloteau99} found that the $^{13}$CO 
emission extends slightly inside the continuum ring in their study of the GG 
Tau A disk; similar data at higher resolution would be valuable.  \hili{These 
dust trap models also predict that larger particles should be concentrated 
at the maximum pressure, although the implied spectral gradient in this 
particular example would not be resolved with our data.  In that sense, pushing 
the angular resolution of multifrequency continuum observations in this and 
other cases should be considered a high priority.}

Although the GG Tau A ring is a particularly spectacular (and useful) example, 
the dust in circumbinary disks more generally could serve as important test 
cases for studying particle traps induced by dynamical interactions with 
companions.  The so-called ``transition" disks, where the companion is 
speculated to be a giant planet, exhibit similar features --- narrow dust 
continuum rings \citep[e.g.,][]{andrews11}, often having more extended gas 
disks with relatively lower dust masses \citep[e.g.,][]{isella12,rosenfeld13b} 
and occasionally showing direct \citep{rosenfeld12,rosenfeld14,casassus13,
vandermarel13,bruderer14} or indirect \citep[e.g.,][]{dong12,follette13} 
evidence for gas well inside those dust rings.  Given that similarity, we 
suggest that there is great value in analyzing resolved multifrequency radio 
observations of dust rings (and gas structures) in both circumbinary and 
transition disks, analogous to the approach presented here.  Ultimately, 
linking these disk targets -- which are undergoing dynamical interactions with 
a broad range of companion masses -- could offer important insights on how 
solids grow inside gas pressure maxima with a wide diversity of strengths and 
shapes.  

\acknowledgments We thank Mark Reid for valuable discussions about data 
modeling \hili{and an anonymous referee for helpful suggestions}.  S.~M.~A.~and 
T.~B.~acknowledge support from NASA Origins of Solar Systems grant NNX12AJ04G.  
A.~I., L.~M.~P., and J.~M.~C.~acknowledge support from NSF award AST-1109334.  
Ongoing CARMA development and operations are supported by the National Science 
Foundation under a cooperative agreement, and by the CARMA partner 
universities.  The VLA is run by the National Radio Astronomy Observatory, a 
facility of the National Science Foundation operated under cooperative 
agreement by Associated Universities, Inc.

\begin{deluxetable}{llcl}
\tablecolumns{4}
\tablewidth{0pt}
\tablecaption{Radio Spectra of GG Tau Components\label{tab:fluxes}}
\tablehead{
\colhead{Component} & \colhead{$\lambda$ (mm)} & \colhead{$S_{\nu}$ (mJy)} & \colhead{Ref.}
}
\startdata
{\bf GG Tau A} & 0.10  &  $5158\pm1410$   & {\it IRAS}                  \\
 & 0.10  &  $6560\pm1469$   & \citet{howard13}            \\
 & 0.14  &  $8995\pm2291$   & {\it AKARI}                 \\
 & 0.15  &  $7620\pm3048$   & \citet{howard13}            \\
 & 0.16  &  $8600\pm3440$   & \citet{howard13}            \\
 & 0.16  &  $6076\pm1974$   & {\it AKARI}                 \\
 & 0.18  &  $8220\pm3288$   & \citet{howard13}            \\
 & 0.19  &  $7130\pm2853$   & \citet{howard13}            \\
 & 0.35  &  $6528\pm1639$   & \citet{aw05}                \\
 & 0.44  &  $4540\pm766$    & \citet{moriarty-schieven97} \\
 & 0.44  &  $4160\pm665$    & \citet{moriarty-schieven97} \\
 & 0.44  &  $2726\pm726$    & \citet{aw05}                \\
 & 0.62  &  $1370\pm382$    & \citet{beckwith91}          \\
 & 0.77  &  $1250\pm323$    & \citet{beckwith91}          \\
 & 0.79  &  $1110\pm163$    & \citet{moriarty-schieven94} \\
 & 0.79  &  $1710\pm176$    & \citet{moriarty-schieven97} \\
 & 0.79  &  $1590\pm170$    & \citet{moriarty-schieven97} \\
 & 0.79  &  $1650\pm183$    & \citet{moriarty-schieven97} \\
 & 0.87  &  $1255\pm138$    & \citet{aw05}                \\
 & 1.06  &  $800\pm206$     & \citet{beckwith91}          \\
 & 1.09  &  $740\pm141$     & \citet{moriarty-schieven94} \\
 & 1.09  &  $1070\pm111$    & \citet{moriarty-schieven97} \\
 & 1.09  &  $830\pm88$      & \citet{moriarty-schieven97} \\
 & 1.09  &  $850\pm90$      & \citet{moriarty-schieven97} \\
 & 1.12  &  $770\pm78$      & \citet{pietu11}             \\
 & 1.25  &  $593\pm130$     & \citet{beckwith90}          \\
 & 1.26  &  $690\pm75$      & \citet{moriarty-schieven97} \\
 & 1.26  &  $630\pm66$      & \citet{moriarty-schieven97} \\
 & 1.26  &  $630\pm70$      & \citet{moriarty-schieven97} \\
 & 1.31  &  $558\pm58$      & this paper                  \\
 & 1.33  &  $557\pm56$      & \citet{harris12}            \\
 & 1.40  &  $604\pm61$      & \citet{guilloteau99}        \\
 & 1.92  &  $320\pm68$      & \citet{moriarty-schieven97} \\
 & 2.68  &  $85\pm10$       & \citet{dutrey94}            \\
 & 2.78  &  $73\pm15$       & \citet{looney00}            \\
 & 2.83  &  $79\pm9$        & this paper                  \\
 & 3.06  &  $41\pm9$        & \citet{ohashi91}            \\
 & 3.40  &  $38\pm4$        & \citet{guilloteau99}        \\
 & 6.92  &  $3.24\pm0.77$   & \citet{rodmann06}           \\
 & 7.29  &  $2.67\pm0.29$   & this paper                  \\
 & 19.09 &  $0.25\pm0.05$   & \citet{scaife13}            \\
 & 50.00 &  $0.036\pm0.007$ & this paper                  \\
\hline
{\bf GG Tau Ba} & 1.31  &  $<9$            & this paper                  \\
 & 1.33  &  $<7$            & \citet{harris12}            \\
 & 2.83  &  $<1.7$          & this paper                  \\
 & 7.29  &  $0.058\pm0.014$ & this paper                  \\
 & 50.00 &  $<0.02$         & this paper                  \\
\hline
{\bf GG Tau Bb} & 1.31  &  $<9$            & this paper                  \\
 & 1.33  &  $<7$            & \citet{harris12}            \\
 & 2.83  &  $<1.7$          & this paper                  \\
 & 7.29  &  $<0.04$         & this paper                  \\
 & 50.00 &  $<0.02$         & this paper                  \\
\hline
{\bf GG Tau/N} & 1.31  &  $<14$           & this paper                  \\
 & 2.83  &  $<2.0$          & this paper                  \\
 & 7.29  &  $0.71\pm0.07$   & this paper                  \\
 & 19.09 &  $2.23\pm0.12$   & \citet{scaife13}            \\
 & 20.03 &  $3\pm1$         & \citet{bieging84}           \\
 & 50.00 &  $2.84\pm0.14$   & this paper                  \\
 & 61.37 &  $3.7\pm0.4$     & \citet{bieging84}           \\
\enddata
\tablecomments{The uncertainties on the flux densities include both statistical 
and systematic calibration terms (added in quadrature).  Upper limits are 
quoted at the 3\,$\sigma$ level, assuming point source emission.  The GG Tau B 
and N flux densities were determined after correction for the primary beam 
responses in each observation.}
\end{deluxetable}

\clearpage

\begin{deluxetable}{lccc}
\tablecolumns{4}
\tablewidth{0pt}
\tablecaption{Inferred Visibility Model Parameters \label{tab:model}}
\tablehead{
\colhead{Parameter} & \colhead{1.3\,mm} & \colhead{2.8\,mm} & \colhead{7.3\,mm}}
\startdata
$S_{\nu,r}$ (mJy)           & $543\pm21$       & $77\pm4$              & $2.23\,^{+0.08}_{-0.12}$ \\
$\mu_r$ (AU)                & $232\pm3$        & $235\pm3$             & $234\pm3$ \\
$\sigma_r$ (AU)             & $26\pm5$         & $29\,^{+4}_{-5}$      & $17\,^{+4}_{-8}$ \\
$i$ (\degr)                 & $37\pm1$         & $37\pm1$              & $37\pm1$ \\
$\varphi$ (\degr)           & $7\pm2$          & $7\pm2$               & $7\pm2$ \\
$S_{\nu,c}$ (mJy)           & $15\,^{+3}_{-7}$ & $2.2\,^{+0.6}_{-0.9}$ & $0.44\,^{+0.02}_{-0.04}$ \\
$\Delta \alpha_c$ (\arcsec) & $+0.03\pm0.04$   & $+0.06\pm0.05$        & $+0.05\pm0.04$ \\
$\Delta \delta_c$ (\arcsec) & $-0.10\pm0.05$   & $-0.09\pm0.06$        & $-0.08\pm0.03$ 
\enddata
\tablecomments{The quoted uncertainties correspond to 68\%\ ($\sim$1\,$\sigma$) 
confidence intervals; the associated $\sim$10\%\ calibration uncertainty is not 
applied to the flux density parameters.  The nuisance parameters describing the 
offset of the ring center from the observed phase center are not included.}
\end{deluxetable}

\clearpage

\begin{deluxetable}{lc}
\tablecolumns{2}
\tablewidth{0pt}
\tablecaption{Inferred Spectrum Model Parameters \label{tab:spec}}
\tablehead{
\colhead{Parameter} & \colhead{Best-Fit Value ($\pm$1\,$\sigma$)}}
\startdata
$\Sigma_0$ (g cm$^{-2}$)          & $0.02\pm0.01$          \\
$\mu_r$ (AU)                      & ($235\pm5$)            \\
$\sigma_r$ (AU)                   & ($25\pm5$)             \\
$T_r$ (K)                         & $25\pm5$               \\
$a_{\rm max}$ (mm)                & $1.5\,^{+0.3}_{-0.9}$  \\
$q$                               & $2.4\,^{+0.3}_{-1.0}$  \\
$\log{S_{\nu,0}^{\rm dust}}$ (Jy) & (-$4.65\pm0.20$)       \\
$\alpha^{\rm dust}$               & ($2.1\pm0.2$)          \\
$\log{S_{\nu,0}^{\rm ff}}$ (Jy)   & (-$4.50\pm0.20$)       \\
$\alpha^{\rm ff}$                 & ($0.0\pm0.1$)          \\
\enddata
\tablecomments{These parameters are valid for the \citet{ricci10a} dust mixture 
with 30\%\ porosity.  See the text for a discussion of alternative 
assumptions.  The values in parenthesis reflect the Gaussian priors we assumed 
for these fits.  The formal reduced $\chi^2$ statistic for the best-fit model 
is $\sim$1.8 (see text).}
\end{deluxetable}

\clearpage

\bibliography{references}
\end{document}